\documentclass[aps,
prb,
preprintnumbers,
superscriptaddress,
footinbib,
amsfonts,
amssymb,
amsmath,
intlimits,
twocolumn,
]{revtex4-2}
\usepackage[utf8]{inputenc}
\usepackage[T2A]{fontenc}
\usepackage{bm,latexsym,mathrsfs,enumerate,amsmath,amssymb}
\usepackage[breaklinks=true,
unicode=true,
urlcolor = blue,
colorlinks = true,
citecolor = blue,
linkcolor = blue
]{hyperref}
\usepackage{graphicx}
\renewcommand{\vec}[1]{\bm{#1}}

\usepackage{chemformula}

\usepackage[final]{pdfpages}

\makeatletter
\AtBeginDocument{\let\LS@rot\@undefined}
\makeatother

\begin{document}
	
\title{Antiferromagnetic nanoscale bit arrays of magnetoelectric \ch{Cr2O3} thin films}
	
\author{Peter Rickhaus}
\email{peter.rickhaus@qnami.ch}
\affiliation{Qnami AG, Hofackerstrasse 40 B, CH-4132 Muttenz, Switzerland}

\author{Oleksandr~V.~Pylypovskyi}
\email{o.pylypovskyi@hzdr.de}
\affiliation{Helmholtz-Zentrum Dresden-Rossendorf e.V., Institute of Ion Beam Physics and Materials Research, 01328 Dresden, Germany}
\affiliation{Kyiv Academic University, Kyiv 03142, Ukraine}

\author{Gediminas Seniutinas}
%\email{gediminas.seniutinas@qnami.ch}
\affiliation{Qnami AG, Hofackerstrasse 40 B, CH-4132 Muttenz, Switzerland}

\author{Vicent~Borras}
%\email{vicent.borras@qnami.ch}
\affiliation{Qnami AG, Hofackerstrasse 40 B, CH-4132 Muttenz, Switzerland}

\author{Paul Lehmann}
\affiliation{Department of Physics, University of Basel, Klingelbergstrasse 82, Basel CH-4056, Switzerland}

\author{Kai Wagner}
\affiliation{Department of Physics, University of Basel, Klingelbergstrasse 82, Basel CH-4056, Switzerland}

\author{Liza \v{Z}aper}
\affiliation{Qnami AG, Hofackerstrasse 40 B, CH-4132 Muttenz, Switzerland}
\affiliation{Department of Physics, University of Basel, Klingelbergstrasse 82, Basel CH-4056, Switzerland}

\author{Paulina J. Prusik}
\affiliation{Helmholtz-Zentrum Dresden-Rossendorf e.V., Institute of Ion Beam Physics and Materials Research, 01328 Dresden, Germany}

\author{Pavlo~Makushko}
\affiliation{Helmholtz-Zentrum Dresden-Rossendorf e.V., Institute of Ion Beam Physics and Materials Research, 01328 Dresden, Germany}

\author{Igor Veremchuk}
\affiliation{Helmholtz-Zentrum Dresden-Rossendorf e.V., Institute of Ion Beam Physics and Materials Research, 01328 Dresden, Germany}

\author{Tobias Kosub}
\affiliation{Helmholtz-Zentrum Dresden-Rossendorf e.V., Institute of Ion Beam Physics and Materials Research, 01328 Dresden, Germany}

\author{Ren\'{e} H\"{u}bner}
\affiliation{Helmholtz-Zentrum Dresden-Rossendorf e.V., Institute of Ion Beam Physics and Materials Research, 01328 Dresden, Germany}
% ORCID 0000-0002-5200-6928

\author{Denis~D.~Sheka}
\affiliation{Taras Shevchenko National University of Kyiv, 01601 Kyiv, Ukraine}

\author{Patrick Maletinsky}
\affiliation{Department of Physics, University of Basel, Klingelbergstrasse 82, Basel CH-4056, Switzerland}

\author{Denys~Makarov}
\email{d.makarov@hzdr.de}
\affiliation{Helmholtz-Zentrum Dresden-Rossendorf e.V., Institute of Ion Beam Physics and Materials Research, 01328 Dresden, Germany}
	
\begin{abstract}
    Magnetism of oxide antiferromagnets (AFMs) has been studied in single crystals and extended thin films. The properties of AFM nanostructures still remain underexplored. Here, we report on the fabrication and magnetic imaging of granular 100-nm-thick magnetoelectric \ch{Cr2O3} films patterned in circular bits with diameters ranging from 500 down to 100\,nm. With the change of the lateral size, the domain structure evolves from a multidomain state for larger bits to a single domain state for the smallest bits. Based on spin-lattice simulations, we show that the physics of the domain pattern formation in granular AFM bits is primarily determined by the energy dissipation upon cooling, which results in motion and expelling of AFM domain walls of the bit. Our results provide a way towards the fabrication of single domain AFM-bit-patterned memory devices and the exploration of the interplay between AFM nanostructures and their geometric shape.

    % 150 words at max
\end{abstract}

\maketitle

\section{Introduction}

Antiferromagnetic spintronics offers major advantages over their ferromagnetic counterpart in terms of stability and density of stored information and operation speed. Recent fundamental explorations include ultrafast processes in spintronics and magnonics \cite{Li20c,Yuan18b,Li20d,Baltz24} with strong focus on the physics of antiferromagnetic (AFM) materials targeting the understanding of the origin of chiral interactions~\cite{Qaiumzadeh18a}, spin-orbit~\cite{Otxoa20,Xue21} and spin-transfer torques~\cite{Ghosh22}, topological features in momentum and real space~\cite{Smejkal18,Gomonay18}, finite size effects~\cite{He12a,Pati16}, strain effects~\cite{Kota17a,Mahmood21,Makushko22}, and the interaction of topologically nontrivial magnetic textures with lattice defects~\cite{Kalita04,Galkina20,Reimers22,Wittmann22,Meer23a,Tan24}. There are numerous application-relevant demonstrations with AFM materials. Available proposals of AFM devices already include magnetoelectric spin–orbit (MESO) logic~\cite{Manipatruni19,Pham20,Prasad20,Vaz21,Vaz24,Fert24}, AFM random access memory (RAM)~\cite{Wadley16,Olejnik17,Reimers23a}, magnetoelectric RAM~\cite{He10a,Kosub17}, and domain-wall-based memory devices~\cite{Hedrich21}. 

By now, research was focused primarily on extended AFM thin films and micro-patterned elements of different families of AFM materials including conducting antiferromagnets with bulk N\'{e}el spin-orbit torques like \ch{Mn2Au} and CuMnAs~\cite{Reimers23}, or simple oxides like NiO~\cite{Meer22}, \ch{Fe2O3}~\cite{Wittmann22,Jani21a,Tan24}, and \ch{Cr2O3}~\cite{Makushko22,Wang22c,Ujimoto23}. Those studies, which are performed on single crystals or extended thin films, are important for fundamental explorations, especially for domain wall physics and switching of the order parameter by external means. To explore the full application potential of these antiferromagnets, their behavior when patterned down to nanoscale dimensions must be known. To this end, there is a strong inspiration from the community working on complex oxides, including \ch{LaFeO3}~\cite{Folven10,Bang19} and \ch{BiFeO3}~\cite{Manipatruni19}. In particular, detailed characterization of sub-$\mu$m \ch{BiFeO3} samples~\cite{Johann11,Steffes19} resulted in the realization of the MESO concept~\cite{Manipatruni19}, which is considered promising for prospective low-energy logic devices. 

In the family of insulating AFMs, magnetoelectric \ch{Cr2O3} attracted attention due to the  possibility to manipulate the magnetic order parameter magnetoelectrically~\cite{Ashida14,Kosub17,Shiratsuchi21} and even by electric fields~\cite{Mahmood21} or spin-orbit torques only~\cite{He24a}, which is paving the way towards AFM magnetoelectric RAM~\cite{Liang21}. Hence, being inspired by initial works on single crystals~\cite{He10a}, there is active exploration of the performance of \ch{Cr2O3} thin films recently, extending studies to unconventional substrates like mica~\cite{Lai23}. It is established that thin films of \ch{Cr2O3} reveal flexomagnetic effects~\cite{Makushko22} and feature finite-size effects for ultrathin films, reflected in the reduction of the transition temperature~\cite{He12a}. To assess the technological relevance of \ch{Cr2O3} for high-areal-density magnetoelectronics, it is important to understand the physics of magnetic states in sub-$\mu$m bits of \ch{Cr2O3} thin films. Furthermore, a robust method to read out the magnetic state of nanoscale bits of AFM thin films should be established. 

\begin{figure*}
	\centering
	\includegraphics[width=\linewidth]{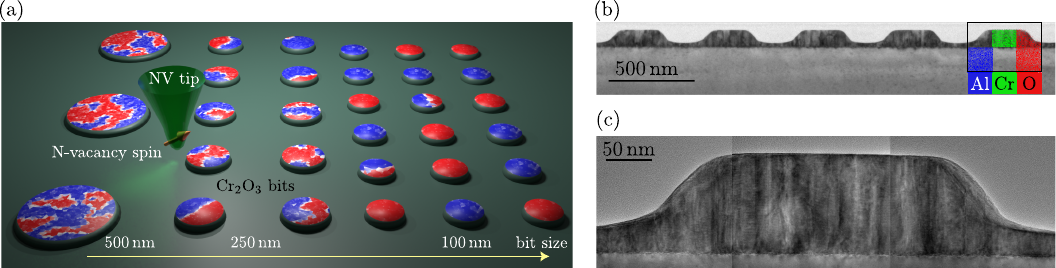}
	\caption{\textbf{Nanoscale arrays of antiferromagnetic bits.} (a)~Schematics of the experimental setup, showing granular bits with AFM domain patterns indicated in red-blue color. The image shows that with the reduction of the bit diameter, there is a change in the domain pattern from multi- to single-domain. (b)~Cross-section transmission electron microscopy image showing a regular array of \ch{Cr2O3} bits with a nominal diameter of 250 nm. The inset depicts the element distribution maps from spectrum imaging analysis based on energy-dispersive X-ray spectroscopy (EDXS). (c) Stitched TEM images showing the granular structure of an individual \ch{Cr2O3} bit with a lateral grain size in the range of 20\,nm.}
	\label{fig:characterization}
\end{figure*}

Here, we demonstrate the fabrication and measurements of nanoscale AFM \ch{Cr2O3} bits, which can accommodate an information bit-stored in the AFM order parameter. By using scanning nitrogen vacancy magnetometry  (SNVM), which has proven useful to read out ferromagnetic magnetic random access memory (MRAM) bits~\cite{Borras24}, we experimentally observe that reducing the lateral size of individual AFM bits with 20-nm-sized grains in an array from 500\,nm down to 100\,nm, there is a transition from multidomain to single domain state. The physics of multidomain states in uniaxial AFM thin films of \ch{Cr2O3} is attributed to the pinning of AFM domain walls at grain boundaries. We performed spin lattice simulations and determined the relevant inter-grain coupling parameters of the order of 15\% of the nominal value with a wide distribution of the inter-grain exchange bonds, which allows matching the experimental and theoretically calculated AFM domain states. This is the first demonstration that \ch{Cr2O3} thin films can be used to realize nanoscale bit-patterned media. To this end, we show that \ch{Cr2O3} can be used to realize arrays of bits with a diameter of 100\,nm and a period of 200\,nm. Imaging of the magnetic states in the 100-nm bits of  thin-film AFM \ch{Cr2O3} reveals their single domain state, as confirmed with SNVM imaging from which the map of the N\'{e}el vector is inferred using machine learning algorithms. 

\section{Results}

\subsection{AFM bit arrays in \ch{Cr2O3} thin films}

We deposited 200-nm-thick \ch{Cr2O3} films by reactive evaporation on $c$-cut sapphire substrates. Magnetotransport characterizations relying on the zero-offset Hall measurement scheme~\cite{tensometer,Kosub17} reveal the N\'{e}el temperature of the as-prepared samples to be about 301\,K~\cite{Kosub15}. The magnetic length in \ch{Cr2O3} is $\ell \approx 20$\,nm~\cite{Hedrich21}. By using electron beam lithography and reactive etching, thin films are patterned in square arrays of circular bits with diameters of 500\,nm (period: 1000\,nm), 250\,nm (period: 500\,nm), and 100\,nm (period: 200\,nm), see Fig.~\ref{fig:characterization}a. Etching of thin films to realize bits leads to a reduction of the film thickness down to about 100\,nm, as confirmed by transmission electron microscopy (TEM) imaging (Fig.~\ref{fig:characterization}b,c). Furthermore, high-resolution TEM analysis provides access to the granular morphology of the thin films with grain size of about 20\,nm and a high crystallinity within each grain (Fig.~\ref{fig:characterization}c). We note that individual bits are connected with a 20-nm-thick \ch{Cr2O3} layer. As \ch{Cr2O3} with a thickness of less than 30\,nm is paramagnetic at room temperature~\cite{Sahoo07,He12a}, these bridges do not affect the interpretation of the SNVM contrast and do not contribute to the measured stray fields, as confirmed by the SNVM measurements. 

\subsection{Imaging of magnetic states of \ch{Cr2O3} bit arrays}

\begin{figure*}
	\centering
	\includegraphics[width=\linewidth]{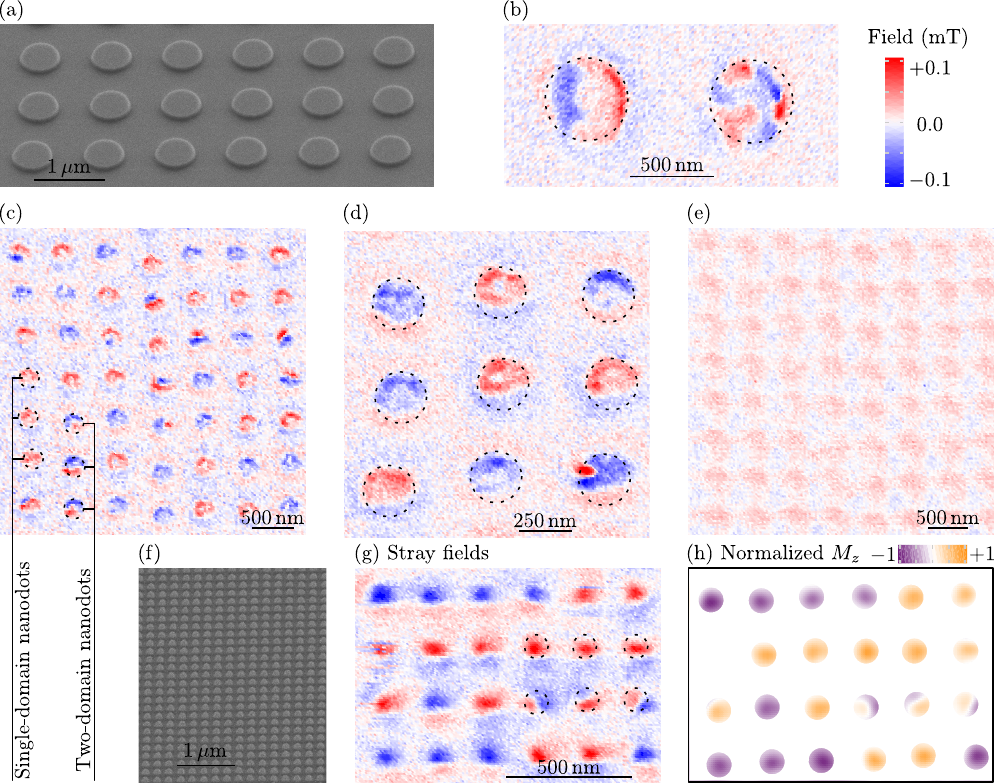}
	\caption{\textbf{Stray field maps and domain structure of AFM bits.} (a)~Scanning electron microscopy (SEM) image of bits (500\,nm in diameter) and (b)~respective SNVM scan of two bits. Here, and in other panels, the dashed lines are guides to the eye and depict the selected bit position. The color bar shows the change of the stray field in the range between $-0.1$ and $+0.1$\,mT. (c,d)~Array of bits (250\,nm in diameter) in magnetically disordered (after ZFC) and (e)~ordered (after FC) states. Dashed lines show exemplary bits which are in single- or two-domain states. (f)~SEM image of the bits (100\,nm in diameter) and (g)~respective SNVM scans. The dashed circles are guides to the eye for three single- and three two-domain bits. (h)~Reconstruction of the out-of-plane magnetization from stray fields shown at the positions of the bits corresponding to panel~(g). The background is blanked as a guide to the eye.}
	\label{fig:nv-scans}
\end{figure*}

We image the magnetic states of the bits in the arrays via SNVM~\cite{Appel19, Hedrich21,Borras24}. Measurements are performed on a commercialy available SNVM system (Qnami ProteusQ). The magnetic state of the samples is prepared by their annealing above the N\'{e}el temperature up to $90^\circ$C and cooling down without magneto-electric field (zero-field cooling, ZFC procedure) or in an applied magnetic field of $B=550$\,mT and an electric field of $E = 1.4$\,MV/m (field cooling, FC procedure). Field cooling allows us to prepare the sample with the N\'{e}el vector having a preferential direction. In contrast, the ZFC procedure results in a sample without a preferred orientation of the N\'{e}el vector. First, we studied the largest bits with a diameter of 500\,nm (Fig.~\ref{fig:nv-scans}a). The map of the magnetic stray fields $\vec{B}_\textsc{nv}$ of two 500-nm bits is shown in Fig.~\ref{fig:nv-scans}b. The magnetic contrast per bit indicates that \textit{all} measured bits of that size are in a multidomain state (see Supplementary Fig.~1), where the latter is identified by the magnetic stray field, $B_\textsc{nv}$ showing a nonzero number of sign reversals (i.e. changes from red to blue imaging contrast) across the extension of the bit.

With a reduction of the bit diameter to 250\,nm, some bits are still found in a two-domain state after the ZFC procedure (Fig.~\ref{fig:nv-scans}c,d). We note that the number of bits with positive vs negative stray field contrast instead of the uniform one is about 30\% (28~of 101~measured bits), which is in line with the assumption that the magnetic state is prepared by thermal demagnetization. Furthermore, we observe no specific pattern in the magnetic state of the bits in the array (Fig.~\ref{fig:nv-scans}c and Supplementary Fig.~2), which suggests that bits are magnetically decoupled. Unlike the signal from the bits, the signal between them is zero on average. The magnetic state of the bit array can be changed by magnetoelectric field cooling the sample from a temperature above the N\'{e}el temperature. A representative image of the ordered magnetic state of 200-nm-diameter bits is shown in Fig.~\ref{fig:nv-scans}e. In this case, the out-of-plane magnetoelectric cooling selects a preferred orientation of the N\'{e}el vector, rendering all bits to be in the same magnetic state.

Lithographically, it is possible to fabricate magnetic bits in \ch{Cr2O3} with smaller diameters. We fabricated extended bit arrays containing bits with a diameter of 100\,nm and a period of 200\,nm (Fig.~\ref{fig:nv-scans}f-h). Using SNVM, we are able to detect signals from individual bits in the array (Fig.~\ref{fig:nv-scans}g and Supplementary Fig.~3). Using the map of magnetic stray fields, we performed a reconstruction of the magnetic moment (Fig.~\ref{fig:nv-scans}h). The reconstruction is done under the assumption that the magnetic texture is homogeneous along the thickness of the bit and the N\'{e}el vector is pointing along the $c$-axis. With the information that stray field maps are measured in a parallel plane at the known SNVM tip height, the magnetization can be reconstructed relying on artificial intelligence algorithms~\cite{Thomas96,Lima09,Casola18,Dubois22,Pylypovskyi23i}. The magnetic pattern shown in Fig.~\ref{fig:nv-scans}h corresponds to the thermally demagnetized state. We found only about 6\% of bits which may be in a multidomain state (9~of 157~measured bits). Three bits in a two-domain state are shown in Fig.~\ref{fig:nv-scans}g.

\subsection{Magnetic states in nanoscale bits of \ch{Cr2O3}}

\ch{Cr2O3} behaves as a two-sublattice collinear antiferromagnet. The macroscopic magnetic state of \ch{Cr2O3} is described by the N\'{e}el vector (primary order parameter) $\vec{n} = 0.5(\vec{M}_1 - \vec{M}_2)$ defined as the difference of the unit vectors of the magnetization of the sublattices $\vec{M}_1$ and $\vec{M}_2$~\cite{Dzialoshinskii57}. For the case of $c$-plane samples, the direction of $\vec{n}$ can be associated with the direction of the boundary magnetization at the top surface~\cite{Andreev96,Belashchenko10}. Following the approach~\cite{Pylypovskyi23i}, we modelled individual bits as granular media. In simulations, we represent each bit as a two-dimensional AFM bipartite square lattice whose dynamics is governed by the Landau--Lifshitz--Gilbert equation. Limiting the consideration of effects stemming from the exchange stiffness and anisotropy only, the static macroscopic magnetic state can be formulated within the nonlinear $\sigma$-model as for \ch{Cr2O3}~\cite{Hedrich21}. The magnetic sites and associated unit vectors of the magnetic moments $\vec{m}_i$, with $i$ enumerating the spins, are located within a circle corresponding to the bit diameter. The grain structure is modelled via a Voronoi pattern with the average tile size equivalent to 20\,nm, as in the experiment, corresponding to the grain size observed by TEM (Fig.~\ref{fig:characterization}c). To compare experiment and simulations, we calculate the N\'{e}el texture in equilibrium. We vary the strength of exchange bonds $J_\text{b}$ at grain boundaries according to the truncated normal distribution~\cite{Cha13}, with the mean value $j$ and standard deviation $\sigma$ measured in units of the nominal exchange strength $J_\text{g}$. By tailoring the width of the distribution $\sigma$, the coupling between grains can vary from the strictly AFM (narrow distribution) to a mix of antiferro- and ferromagnetically coupled grains within one bit.

\begin{figure*}
	\centering
	\includegraphics[width=\linewidth]{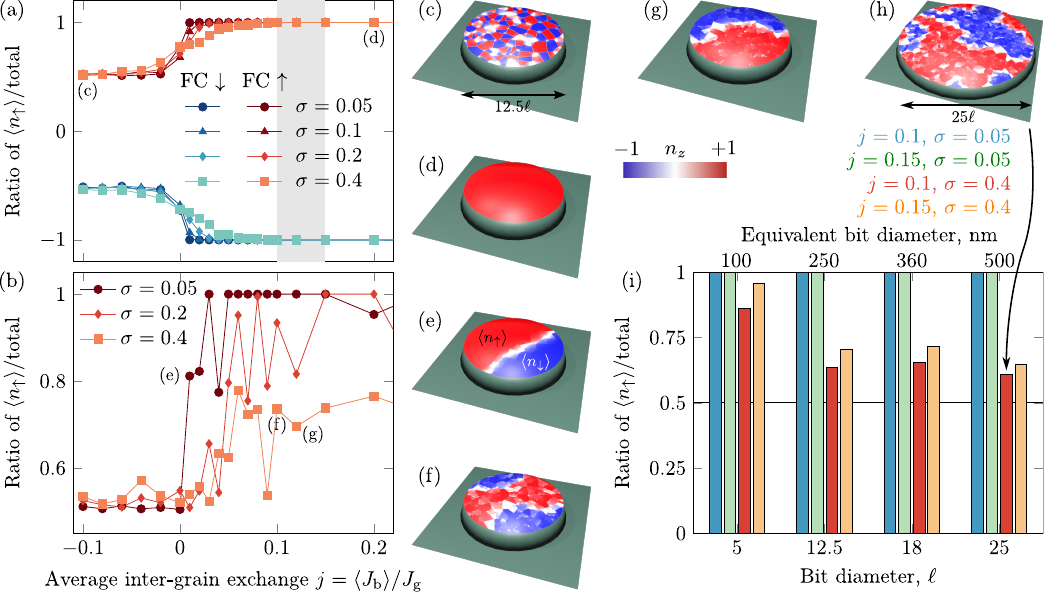}
	\caption{\textbf{Spin-lattice simulations of granular chromia bits.} (a)~FC-like and (b)~ZFC-like simulations for bits with different inter-grain exchange coupling parameters (bit diameter $12.5\ell$). The gray-shaded region shows the range for $j$ determined in~Ref.~\cite{Pylypovskyi23i} for the case of extended thin films with a thickness of 200\,nm. (c)~Magnetic texture in terms of $n_z$ at the bit surface for the case of $j = -0.1$ and~(d)~$j = 0.2$. \mbox{(e--g)}~Examples of the evolution of the domain pattern depending on the inter-grain coupling for the parameters marked in~(b) with the diameter of $12.5\ell$ and (h)~$25\ell$. (i)~Relative area covered by ``up'' domains depending on the size of bits and inter-grain coupling, see detailed statistics in Supplementary Fig.~11.}
	\label{fig:sim-bits}
\end{figure*}

The magnetic state of the bit in simulations can be quantified by the measurement of the related area occupied by the N\'{e}el vectors directed upward, $\langle n_\uparrow \rangle$, in comparison with the total area of the sample. First, we examine the FC-like procedure by relaxing the bit starting from the uniform state in terms of $\vec{n}$, see Fig.~\ref{fig:sim-bits}a. If $j \gtrsim 0.05$, the bit keeps the uniform state for all examined inter-grain coupling parameters (Fig.~\ref{fig:sim-bits}d). Otherwise, for $ 0 < j \lesssim 0.05$, the fraction of oppositely oriented domains grows with an increase of the amount of ferromagnetic exchange bonds. In the limiting case $ j < 0$, a bit tends to be in a frustrated state (Fig.~\ref{fig:sim-bits}c). We note that for the material parameters estimated for similar films with a thickness of 200\,nm~\cite{Pylypovskyi23i}, bits are in the uniform state after the FC procedure, see also~Fig.~\ref{fig:nv-scans}e.

A ZFC-like procedure can be emulated by setting the initial state in simulations to be disordered. For a narrow distribution of exchange bonds ($\sigma = 0.05$), there is a trend for a fast saturation to the uniform state (circular symbols in Fig.~\ref{fig:sim-bits}b). With larger $\sigma$, the multidomain state occurs more frequently for larger $j$ (diamond symbols in Fig.~\ref{fig:sim-bits}b). For a sufficiently large $\sigma \sim 0.4$, a typical state is multidomain (square symbols in Fig.~\ref{fig:sim-bits}b, see exemplary magnetic patterns in Fig.~\ref{fig:sim-bits}e-h). The domain wall structure in these samples is determined by the grain size. Having grains with a characteristic size of the order of the magnetic length, it cannot accommodate the complete domain wall. Thus, the transition between domains consists of almost uniformly magnetized grains with $|n_z|$ being close to 0, see Fig.~\ref{fig:sim-bits}f--h.

Fig.~\ref{fig:sim-bits}i shows the relative area occupied by the domains oriented ``up'' using the ZFC-like procedure. If the distribution of the exchange bonds at grain boundaries is narrow ($\sigma = 0.05$), then the bit can be in the monodomain state independently on its size (blue and green columns in Fig.~\ref{fig:sim-bits}i). In contrast, $\sigma = 0.4$ leads to a reduction of the area of the ``up'' domains towards half of the sample with growth of its size. At the same time, the smallest bits of $5\ell$ in diameter maintain almost the single-domain size independently on $\sigma$.

\begin{figure}[t]
	\centering
	\includegraphics[width=\linewidth]{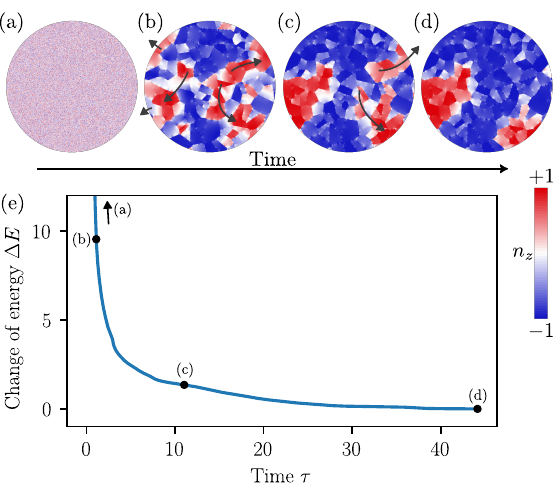}
	\caption{\textbf{Temporal evolution of the domain pattern (bit diameter $12.5\ell$, $j = 0.1$, $\sigma = 0.4$).} (a--d)~Sequential snapshots of the magnetic state for $\omega_0t = 0, 1.1, 11$ and~$44$ (steady state). Dark-gray arrows indicate the direction of the domain wall motion. (e)~Change of the total energy $\Delta E$ measured in units of $J_\text{g}$ with time $\tau$ measured in units of inverse AFM resonance frequency. 
	%and (f)~dynamics rate with time. 
	The black symbols indicate the positions corresponding to panels~(a--d). See Supplementary Fig.~10 for details.
	}
	\label{fig:sim-relaxation}
\end{figure}

The calculations reveal that the described domain wall pinning behavior is a result of an interplay of the energy landscape formed by the grain boundaries and energy surplus formed by the initial paramagnetic state. For a sufficiently small bit, the energy penalty associated with a domain wall is enough to move the domain wall through all the pinning sites at boundaries to reach the uniform state independent of the pinning strength (see Fig.~\ref{fig:sim-relaxation} and Supplementary Figs.~7--9). In larger samples, the surplus of energy is insufficient, and domain walls tend to stop being pinned at grain boundaries and usually touch the sample's boundary. 

\section{Discussion}

To summarize, we experimentally demonstrated a possibility to (i)~realize arrays of AFM \ch{Cr2O3} bits with perpendicular anisotropy, (ii)~perform readout of their magnetic state by means of SNVM, and (iii)~control this state using magnetoelectric cooling. We found that the bit's lateral size is decisive for the spontaneous domain formation during ZFC procedure: while all measured bits of 500\,nm diameter are in a multidomain state, only 6\% of the 100-nm-sized bits are found in a two-domain state (Fig.~\ref{fig:nv-scans}). These results are in a qualitative agreement with spin-lattice simulations: for a broad distribution of exchange bonds characteristic for these type of samples~\cite{Pylypovskyi23i}, smaller samples demonstrate only rare incidents of splitting in a two-domain state due to a weak domain wall pinning at grain boundaries. At the same time, equivalents of 500-nm-sized bits approach a 50/50 ratio between the oppositely oriented AFM domains as a typical state. Spin-lattice simulations show that the smallest bits are most likely to be in a single-domain state. This is in agreement with the domain imaging using SNVM, which identified only several bits in a two-domain state, with the majority being in a single-domain state. 

The variation of the inter-grain magnetic coupling has a strong influence on the domain structure for the $c$-plane cut sample (Fig.~\ref{fig:sim-bits}i). The quality of grain boundaries can be controlled by the fabrication procedure, e.g., annealing temperature, which influences type and distribution of structural and magnetic defects~\cite{Veremchuk22}. We anticipate that our findings will stimulate research on granular antiferromagnets including further miniaturization of AFM bits, consideration of other materials~\cite{Reimers23}, and exploring spin Hall physics at nanoscale islands of \ch{Cr2O3}~\cite{Kosub17}. In particular, the shown interplay between the thermodynamical equilibrium magnetic state and size of the bit opens a way for further studies of the geometric constraints on magnetism in nanosized AFMs with in-plane anisotropy and strain effects~\cite{Gomonay07a,Folven10,Bang19,Lee20a,Meer22} extending them to granular materials and easy-axis anisotropy. Furthermore, this work demonstrates the necessity to improve readout methods for success of using SNVM as a readout method for AFM nanoscale materials. For technological trials, all-electrical-readout for these nanoscale AFM bits should be developed.

\section*{Acknowledgments}

We thank Conrad Schubert (HZDR) for helping with the preparation of \ch{Cr2O3} thin films. This work is supported in part via the German Research Foundation (DFG) under the grants MC 9/22-1, MA 5144/22-1, MA 5144/24-1. Numerical calculations were performed using the Hemera high performance cluster at the HZDR~\cite{hzdrcluster}. Additionally, the use of the HZDR Ion Beam Center TEM facilities and the funding of TEM Talos by the German Federal Ministry of Education and Research (BMBF; grant No. 03SF0451) in the framework of HEMCP are acknowledged.

%\bibliography{qnami-bits}
%apsrev4-2.bst 2019-01-14 (MD) hand-edited version of apsrev4-1.bst
%Control: key (0)
%Control: author (8) initials jnrlst
%Control: editor formatted (1) identically to author
%Control: production of article title (0) allowed
%Control: page (0) single
%Control: year (1) truncated
%Control: production of eprint (0) enabled
%



\section*{Supplementary material}

# Supporting information for "Antiferromagnetic nanoscale bit arrays of magnetoelectric $Cr_2O_3$ thin films"

Peter Rickhaus,[1, *] Oleksandr V. Pylypovskyi,[2, 3, †] Gediminas Seniutinas,[1] Vicent Borras,[1] Paul Lehmann,[4] Kai Wagner,[4] Liza Žaper,[1, 4] Paulina J. Prusik,[2] Pavlo Makushko,[2] Igor Veremchuk,[2] Tobias Kosub,[2] René Hübner,[2] Denis D. Sheka,[5] Patrick Maletinsky,[4] and Denys Makarov[2, ‡]

[1] *Qnami AG, Hofackerstrasse 40 B, CH-4132 Muttenz, Switzerland*
[2] *Helmholtz-Zentrum Dresden-Rossendorf e.V., Institute of Ion Beam Physics and Materials Research, 01328 Dresden, Germany*
[3] *Kyiv Academic University, Kyiv 03142, Ukraine*
[4] *Department of Physics, University of Basel, Klingelbergstrasse 82, Basel CH-4056, Switzerland*
[5] *Taras Shevchenko National University of Kyiv, 01601 Kyiv, Ukraine*

## CONTENTS

## I. CHROMIA THIN FILM FABRICATION AND NANOPATTERNING

We prepared 200-nm-thick $Cr_2O_3$ thin films on *c*-cut single-crystalline $Al_2O_3(0\,0\,0\,1)$ substrates (Crystec GmbH) by reactive evaporation of chromium at 700 °C with a background partial pressure of molecular oxygen of $10^{-5}$ mbar (base pressure: below $10^{-7}$ mbar; deposition rate of about 0.04 nm/s; source-to-sample distance: 60 cm). After the deposition of $Cr_2O_3$, the samples were heated up to 750 °C for several minutes to improve the surface quality of the thin films. Afterwards, the substrate was cooled down by heat dissipation through the sample mounting structure.

## II. TEM CHARACTERIZATION

Cross-sectional bright-field transmission electron microscopy (TEM) and high-resolution TEM (HRTEM) analyses were performed using an image-$C_s$-corrected Titan 80-300 microscope (FEI, Field Electron and Ion Company) operated at an accelerating voltage of 300 kV. High-angle annular dark-field scanning TEM (STEM) imaging and spectrum imaging analysis based on energy-dispersive X-ray spectroscopy (EDXS) were done with a Talos F200X microscope (FEI) operated at 200 kV. Prior to (S)TEM analysis, the specimen was mounted in a double-tilt high-visibility low-background holder and placed for 8 s into a Model 1020 Plasma Cleaner (Fischione) to remove potential contaminations. Cross-sectional preparation of the TEM lamella was done by in situ lift-out using a Helios 5 CX focused ion beam (FIB) device (Thermo Fisher). To protect the sample surface, a carbon cap layer was deposited beginning with electron-beam-assisted and subsequently followed by Ga-FIB-assisted precursor decomposition. Afterwards, the TEM lamella was prepared using a 30-keV Ga-FIB with adapted currents. Its transfer to a 3-post copper lift-out grid (Omniprobe) was done with an EasyLift EX nanomanipulator (Thermo Fisher). To minimize sidewall damage, Ga ions with only 5-keV energy were used for final thinning of the TEM lamella to electron transparency.

---

* peter.rickhaus@qnami.ch
† o.pylypovskyi@hzdr.de
‡ d.makarov@hzdr.de

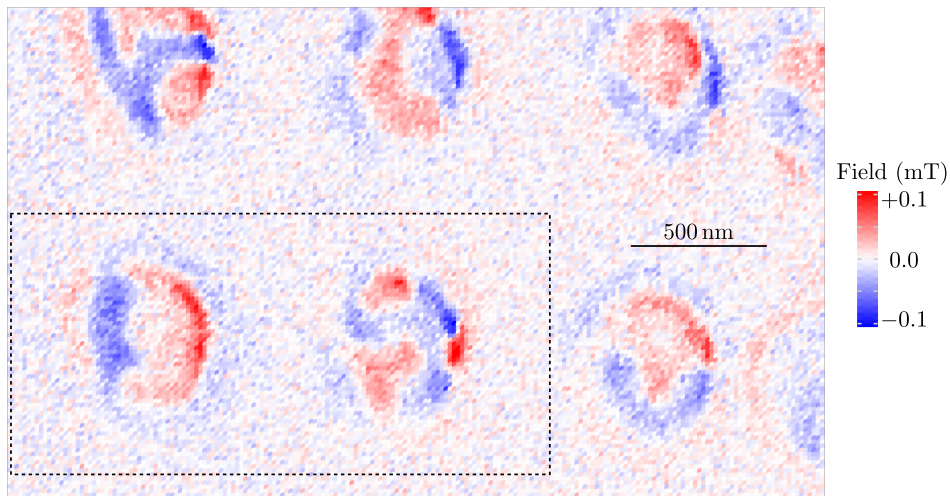

Supplementary Fig. 1. Stray field map for bits of 500 nm in diameter. The dotted area indicates the region shown in Fig. 2b of the main text.

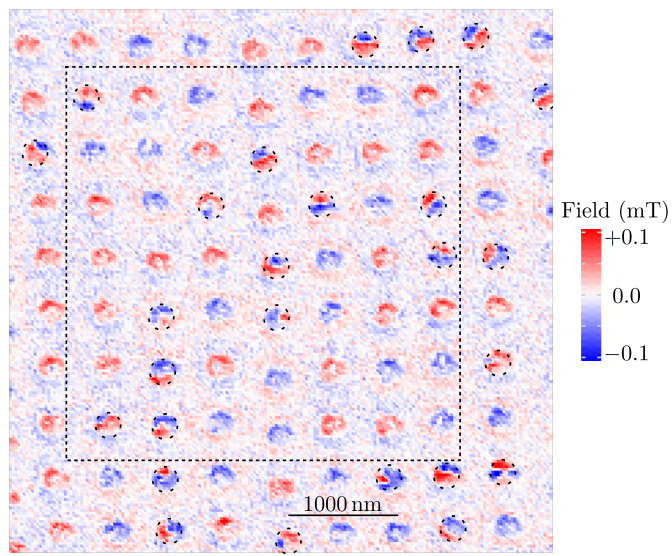

Supplementary Fig. 2. Stray field map for bits of 250 nm in diameter. The rectangular dotted area indicates the region shown in Fig. 2c of the main text. The round dashed lines indicate bits in a multidomain state.

### III. SCANNING NITROGEN VACANCY MAGNETOMETRY

Scanning nitrogen vacancy magnetometry (SNVM) [1] was performed under ambient conditions at about 22° C on the Qnami ProteusQ system. The sensor tips (Qnami Quantilever MX+) are fabricated from single-crystalline $\langle 100 \rangle$-oriented diamond that was implanted with $^{14}$N ions at 12 keV and annealed to form nitrogen vacancy centers. After lithographic fabrication, the resulting diamond scanning probe contains a single nitrogen vacancy center, located at the tip of a parabolic pillar, and provides a spatial resolution of approximately 50 nm [2] and typical sensitivities of approximately $3\,\mu\text{T}/\sqrt{\text{Hz}}$ [2]. Mapping of magnetic stray fields is performed using continuous-wave optically detected magnetic resonance (CW-ODMR) imaging [1]. For a sign-sensitive measurement of stray fields, a small bias magnetic field of about 50 Oe was applied during all measurements. The single-pixel integration time for the presented measurements range from 0.6 s to 7 s.

The reconstruction of the magnetic texture is done under assumption that the magnetic texture is homogeneous along the thickness of the bit and the Néel vector is pointing out-of-plane (parallel to the $c$ axis). With the stray field maps obtained with the nitrogen vacancy center in a parallel plane, magnetization can be reconstructed relying

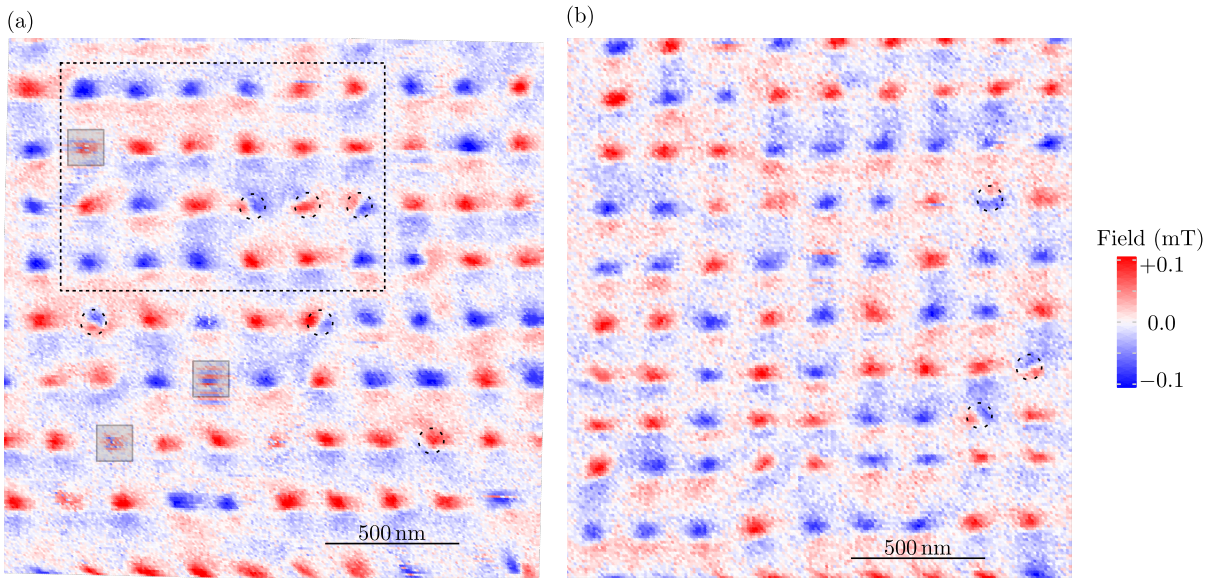

Supplementary Fig. 3. Stray field maps for bits of 100 nm in diameter measured in two different regions on the sample. The rectangular dotted area indicates the region shown in Fig. 2g of the main text. The round dashed lines indicate bits in a multidomain state. The gray semitransparent squares indicate bits within the field of view, which are excluded from consideration as their state cannot be identified.

on the machine learning algorithm presented Dubois *et al.* [3]. In this reconstruction, we iterated over 1000 epochs and assumed the nitrogen vacancy center was at $35 \pm 5$ nm from the sample and with a polar and azimuthal angles of $54.5°$ and $90°$, respectively.

## IV. SPIN-LATTICE SIMULATIONS FOR MANY-GRAIN SAMPLES

In simulations, the magnetic energy of film is given by the anisotropic Heisenberg Hamiltonian

$$\mathcal{H} = -\frac{S^2}{2} \sum_{i,k} J_{ik} \boldsymbol{m}_i \cdot \boldsymbol{m}_k - \frac{KS^2}{2} \sum_i m_{iz}^2, \qquad (1)$$

where $J_{ik}$ is the exchange integral between the $i$-th and $k$-th spins, index $k$ runs over the nearest neighbors of $i$, $S$ is the spin length, $\boldsymbol{m}_i = \{m_{ix}, m_{iy}, m_{iz}\}$ represents the unit vector of the magnetic moment for the $i$-th site and $K$ is the coefficient of the easy axis anisotropy. Equilibrium magnetic textures are addressed by solving the Landau–Lifshitz–Gilibert equation

$$\frac{d\boldsymbol{m}_i}{dt} = \frac{1}{\hbar S} \boldsymbol{m}_i \times \frac{\partial \mathcal{H}}{\partial \boldsymbol{m}_i} + \alpha_\text{G} \boldsymbol{m}_i \times \frac{d\boldsymbol{m}_i}{dt}, \quad i = \overline{1, \mathcal{N}}, \qquad (2)$$

with $\hbar$ being the Planck constant, $\alpha_\text{G}$ being the Gilbert relaxation parameter, and $\mathcal{N}$ being the total number of spins. The set of equations (2) is solved using the spin-lattice simulator SLaSi with graphics processing unit (GPU) acceleration [4]. Following the procedure described in [5], we set $J_{ik} = J_\text{g} = -2.34 \times 10^{-20}$ J as the nominal exchange integral within each grain, $K = 3.656 \times 10^{-24}$ J as the coefficient of the easy axis anisotropy, $S = 1/2$, and $\alpha_\text{G} = 1$. The magnetic length in units of the lattice constant $a$ is $\ell = a\sqrt{|J_\text{g}/K_\text{g}|} = 80a$ to assure a high resolution of the sample's state. The mesh size varies from $400 \times 400$ to $2000 \times 2000$ lattice sites, which are located within a circle of the maximal available diameter corresponding to the size of a magnetic bit. We note that this approach is valid to analyze static distributions of the Néel vector in $Cr_2O_3$ on spatial scales, which are much larger than the size of its crystallographic unit cell. In the main text, we define $E = \mathcal{H}/J_\text{g}$ as the dimensionless energy and $\tau = \omega_0 t$ as the dimensionless time with $\omega_0 = K/(\hbar S)$ being the frequency of antiferromagnetic resonance.

The grain structure is defined by a Voronoi pattern, which determines the exchange bonds that are subjected to changes because of the grain boundary. The inter-grain coupling is simulated as a truncated normal distribution of the exchange bonds $J_{ik}$ with the mean value $j = \langle J_{ik} \rangle / J_\text{g}$ and standard deviation $\sigma = 0.01 \ldots 0.4$ in units of $J_\text{g}$

4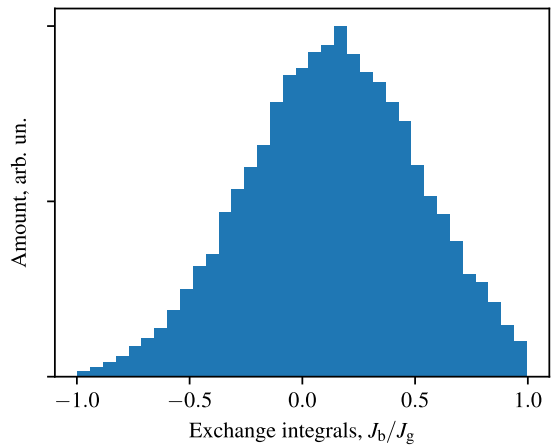

Supplementary Fig. 4. An exemplary distribution of exchange bonds at grain boundaries for $j = 0.15$ and $\sigma = 0.4$.

(Supplementary Fig. 4). For a bipartite square lattice, the antiferromagnetic unit cell is represented by neighboring spins arranged in a square. For the $\nu$-th square with magnetic moments labeled as $\boldsymbol{A}_\nu$, $\boldsymbol{B}_\nu$, $\boldsymbol{C}_\nu$, $\boldsymbol{D}_\nu$ in the counter-clockwise order, the discrete Néel order parameter is defined as $\boldsymbol{n}_\nu = \boldsymbol{A}_\nu - \boldsymbol{B}_\nu + \boldsymbol{C}_\nu - \boldsymbol{D}_\nu$ [6, 7].

Supplementary Fig. 5 shows exemplary results of simulations for the bits with diameters of $5\ell$ and $12.5\ell$ (i.e., equivalent to 100 nm and 250 nm, respectively). The energy drops very fast at the very beginning when the ordering from the random texture to an antiferromagnetic one happens. This is followed with high values of the dynamics rate measured as $\max |d\boldsymbol{m}/dt|$. When it's value drops to be of the order of 1, only domain wall jumps between grains are observed.

Supplementary Fig. 6 shows the evolution of the domain wall structure with the inter-grain coupling. For a weak and more spatially uniform exchange coupling (Supplementary Fig. 6a), the domain wall is almost always represented by the grain boundary itself, while the neighboring grains have the opposite values of $n_z$ close to saturation. In this case, the thickness of the domain walls is primarily determined by the grain size and is about $\ell$. Larger and spatially distributed coupling between grains leads to the appearance of grains which have a substantial amount of spins lying in the hard plane or at least strongly tilted from the easy axis. This state appears because of the average grain size being of about $\ell$, which forces each grain to be in the magnetically uniform state. The average orientation of spins in this case is determined by the magnetic orientation of the surrounding grains. This is especially well pronounced in Supplementary Fig. 6d, where a lot of grains in the vicinity of the domain wall have a weak blue-to-red color gradient indicating value of $n_z$. We note that in case of a narrow distribution of exchange bonds and large enough $j$, the bulk-like domain wall can be pinned, see Supplementary Fig. 6e. The pinning strength of this state on grain boundaries is very small and such textures are rarely observed in simulations.

A possible equilibrium magnetic state of the bit is determined by the combination of its diameter and inter-grain coupling. Exemplary pictures of the temporal evolution in zero-field-cooling-like (ZFC-like) procedure are shown in Supplementary Figs. 7–10. In all cases, the initial state shown in Supplementary Figs. 7a–10a is a random orientation of spins, which are characteristic of a thermally demagnetized bit. A narrow distribution of exchange bonds creates a shallow energy landscape for domain walls. Usually, they do not pin at grain boundaries because of a substantial excess of energy and leave the sample. During their dynamics, the domain wall shape is almost not disturbed by the presence of grains. These states are shown in Supplementary Figs. 7b–f and 8b–l, were the granular structure is not pronounced in the equilibrium magnetic texture but occasionally deforms the domain wall during its movement (Supplementary Fig. 8d). Thus, independent of the bit size, a narrow distribution of exchange bonds leads to a single-domain state, see also Fig. 3i of the main text.

Magnetic dynamics for the bits with a wide distribution of exchange bonds is substantially different. Starting from the random orientation of spins, a small enough sample finally comes to a single-domain state (Supplementary Fig. 9). The Néel texture is represented by almost uniformly ordered grains and domain motion is represented by a reversal of the order parameter within the grain, which is observed for both, small bit of $5\ell$ in diameter (Supplementary Fig. 9b–f) and larger bits of $12.5\ell$ in diameter (Supplementary Fig. 10b–l). The difference in lateral sizes of these two bits influences the energy dissipation process. For a small bit (Supplementary Fig. 9b–l), the initial energy pumped into the system by temperature (the initial random state) is enough to push the domain wall through the whole sample to set it into a single-domain state. Larger bits (Supplementary Fig. 10b–l) have the same areal energy density at the beginning as smaller bits. Therefore, the thermal energy can be insufficient to move the domain wall over a large distance. Hence, the domain wall is affected by pinning at grain boundaries. Finally, this leads to a formation

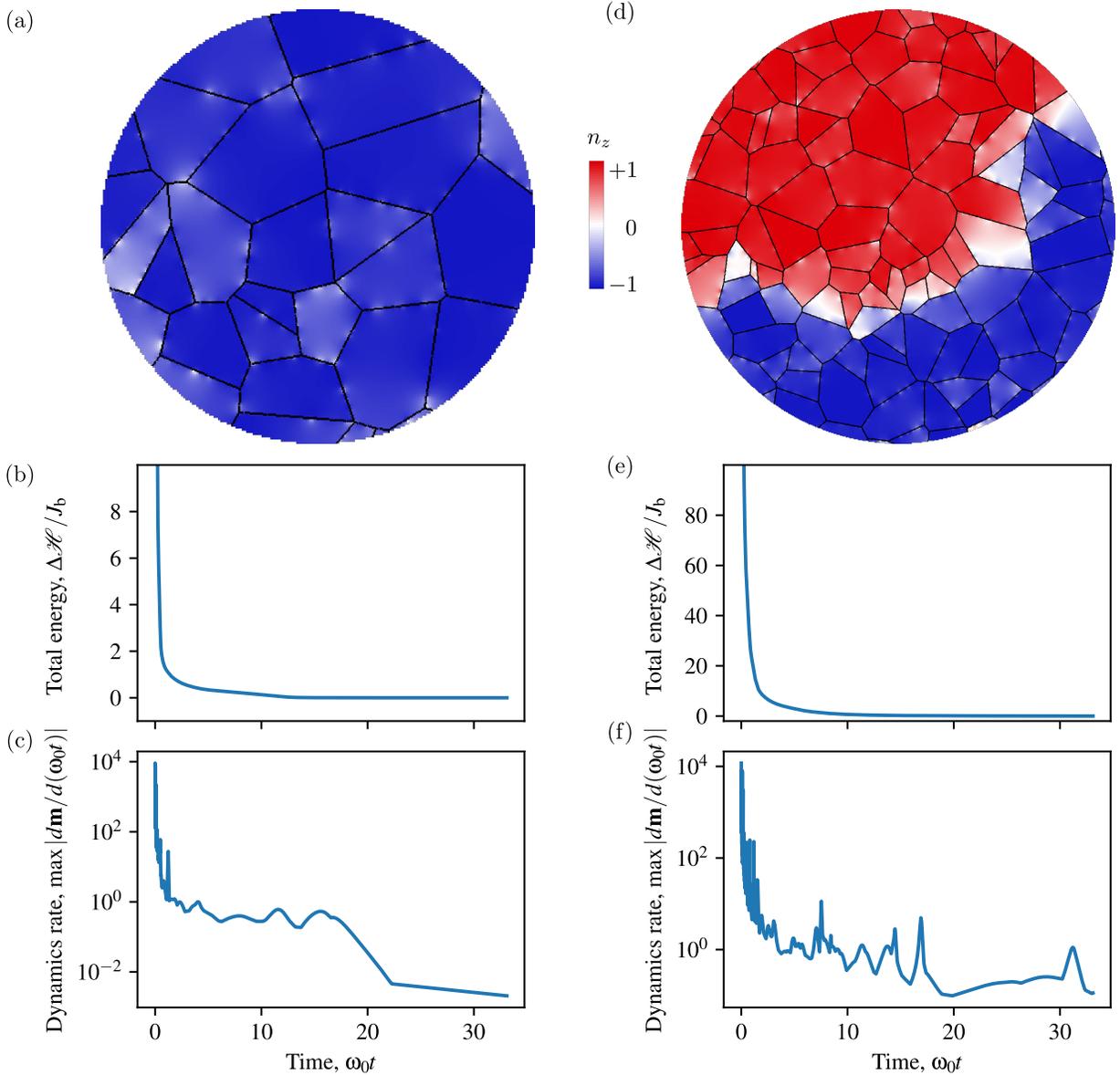

Supplementary Fig. 5. **Simulations of granular bits.** (a) Granular structure of a bit with a diameter of $5\ell$ (100 nm in equivalent) with $j = 0.15$ and $\sigma = 0.4$. The orientation of the Néel vector $\boldsymbol{n}$ is shown with color, grain boundaries are shown by black lines. (b) Change of the energy with time and (c) dynamics rate for the texture shown in panel (a). (d–f) Same for the bit with a diameter of $12.5\ell$ (250 nm in equivalent).

of several large almost uniformly ordered regions contacting with boundaries (Supplementary Fig. 10l). The final dynamics is related to the adjustment of the orientation of $\boldsymbol{n}$ within certain grains.

Fig. 11 summarizes the statistical analysis of the single- or multidomain behavior of granular bits after the ZFC-like procedure extending Fig. 3i in the main text. The data is grouped by different bit diameters corresponding to the samples in experiment. Each dataset presented by a box plot contains 10 simulations started from different random orientations of magnetic moments. All simulations with inter-grain exchange coupling of $j = 0.1$ and $j = 0.15$ with a narrow distribution $\sigma = 0.05$ result in a uniform state independent of the disk diameter, which makes the box plot being degenerated to a line. These points are highlighted by blue and green symbols. A broad distribution of exchange bonds, $\sigma = 0.4$, drastically changes the picture. For a small bit diameter of $5\ell$, a domain wall, if present, is located approximately to the center of the bit. We found two such cases for $j = 0.1$ and none for $j = 0.15$, which is reflected in a tall orange box in Fig. 11 for a bit of the smallest diameter. For bits of lager diameters, the median area occupied by a domain of a certain direction is below 70% and approaches 50% for the largest samples investigated in the simulations. At the same time, only for bits with the diameter of $25\ell$, the single-domain state has not been

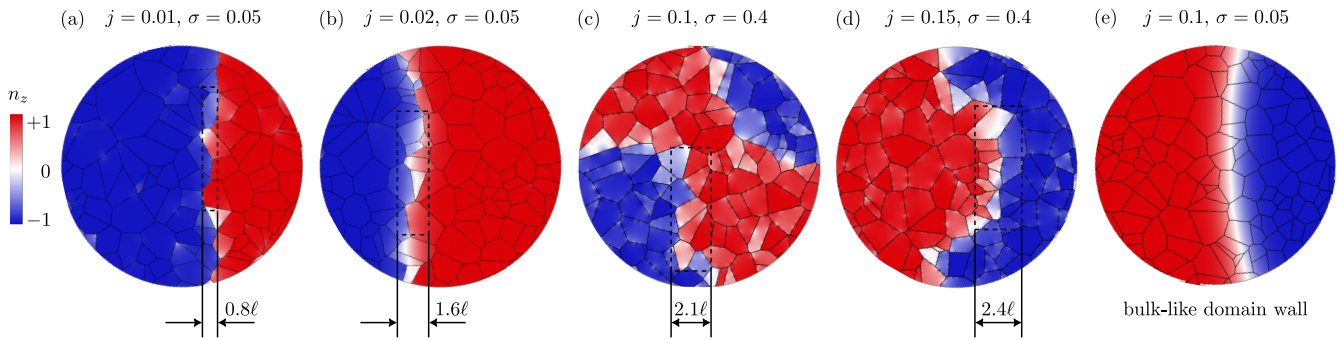

Supplementary Fig. 6. **Domain walls in bits with different inter-grain coupling.** The simulated samples have a diameter of $12.5\ell$. The dashed area shows the domain wall region for which its width is estimated. Grain boundaries are shown by thin black lines. We note that the state (e) is rarely observed in simulations because of a weak pinning for the selected inter-grain coupling.

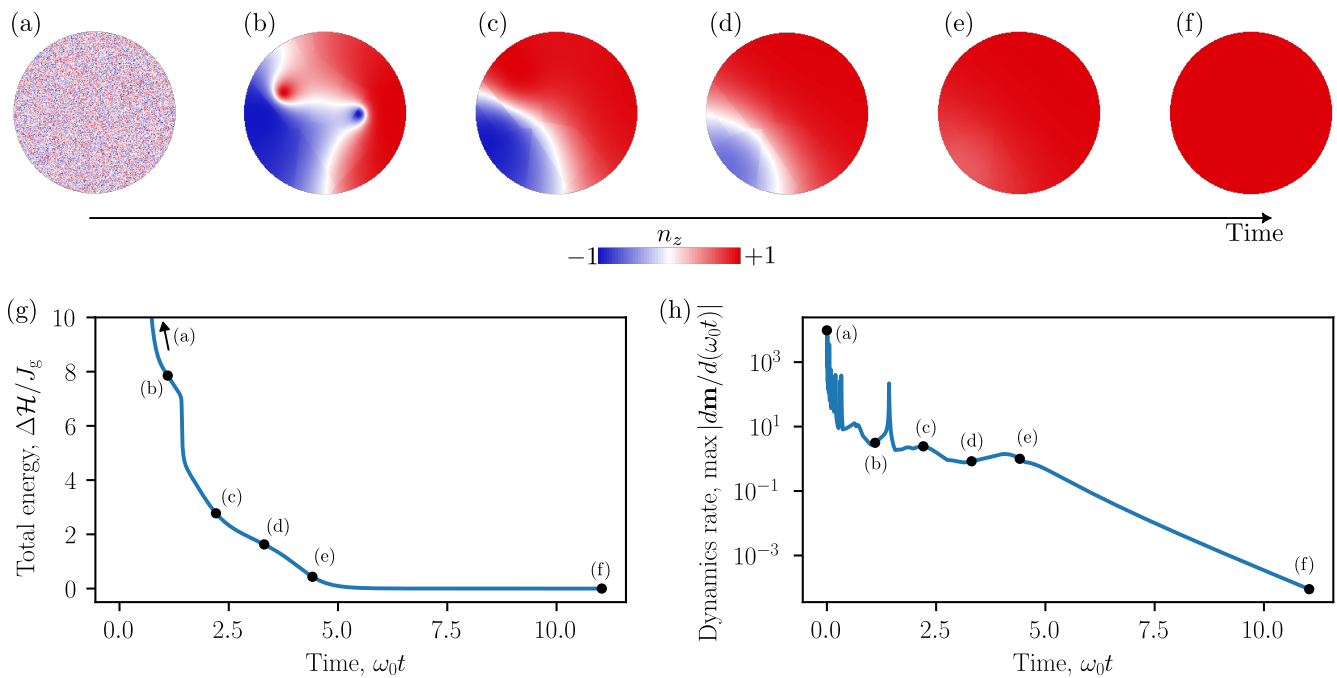

Supplementary Fig. 7. **Magnetic dynamics in ZFC-like simulations (bit diameter $5\ell$, grain coupling is given by $j = 0.1$ and $\sigma = 0.05$).** (a–f) Sequential magnetic states colored according to the value of $n_z$. (g) Change of the total energy with time. The black symbols show the time positions corresponding to panels (a–f). (h) Dynamics rate. Notations are the same as in panel (g).

observed. This is indicated by whiskers ending at the level of about 80% of areal filling of the largest domain; for bits of smaller diameters the top whiskers end almost at the 100% areal filling indicating a possibility to get the single-domain state in the ZFC-like procedure.

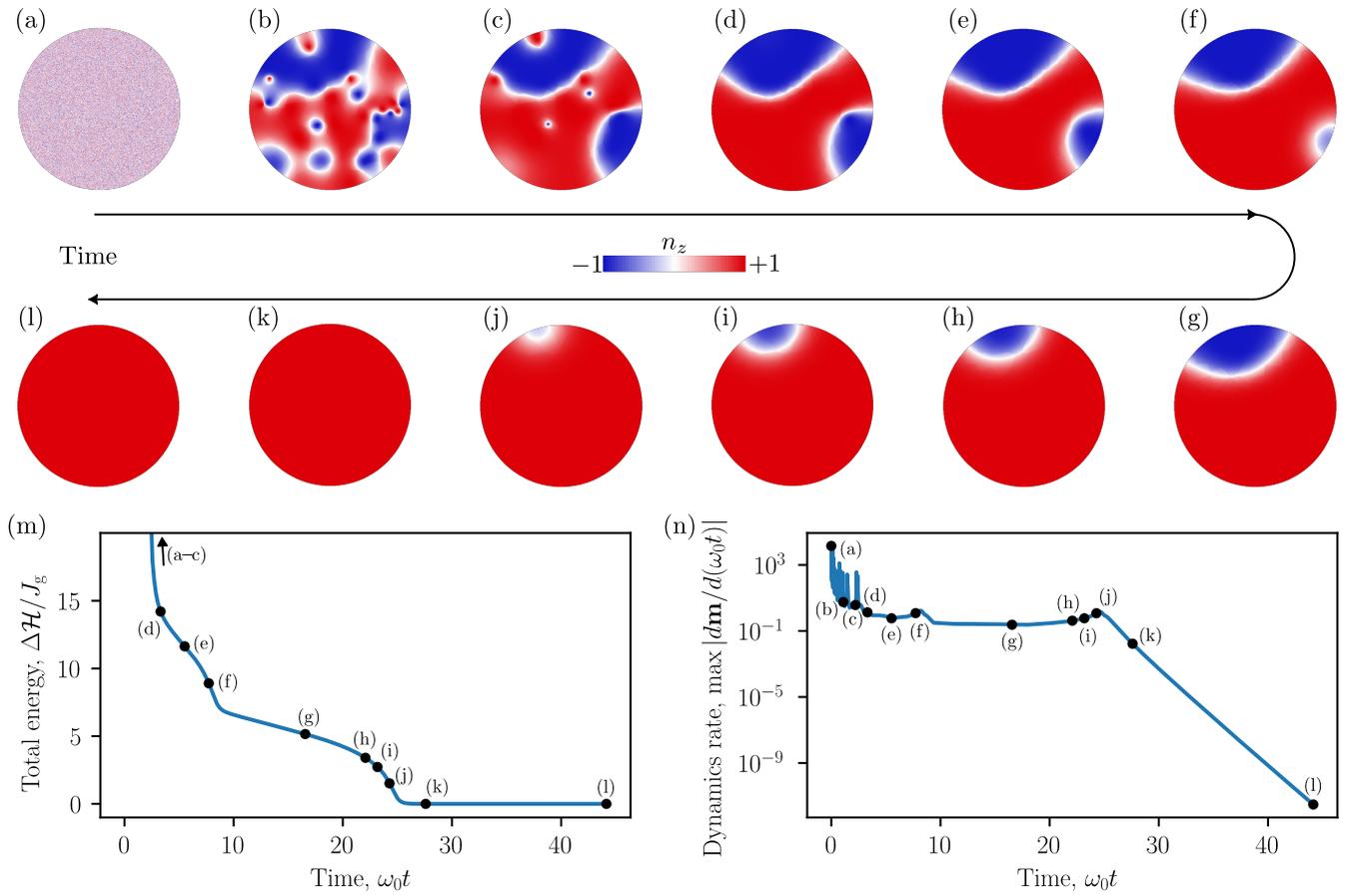

Supplementary Fig. 8. **Magnetic dynamics in ZFC-like simulations (bit diameter $12.5\ell$, grain coupling is given by $j = 0.1$ and $\sigma = 0.05$).** (a–l) Sequential magnetic states colored according to the value of $n_z$. (m) Change of the total energy with time. The black symbols show the time positions corresponding to panels (a–l). (n) Dynamics rate. Notations are the same as in panel (m).

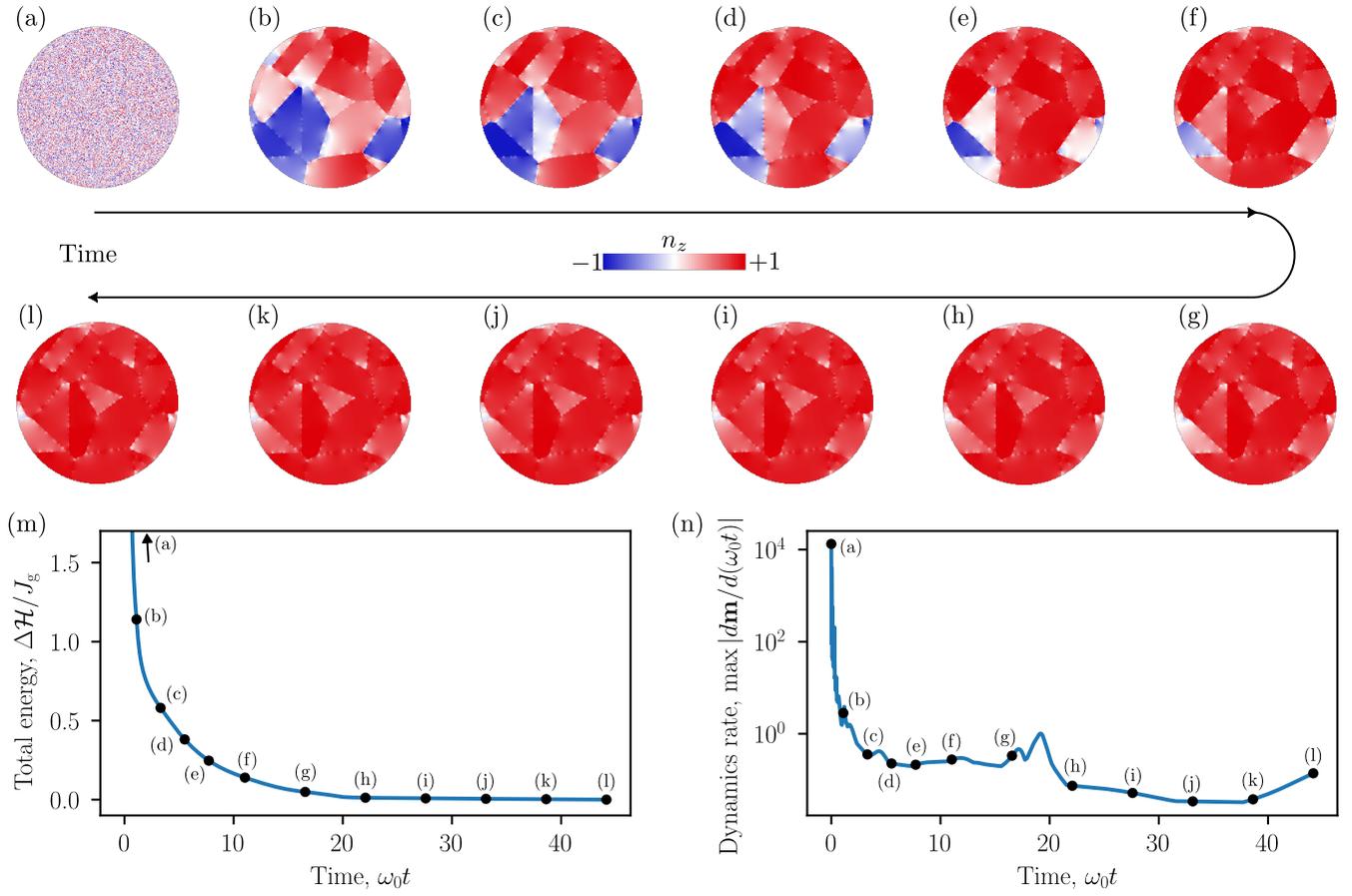

Supplementary Fig. 9. **Magnetic dynamics in ZFC-like simulations (bit diameter $5\ell$, grain coupling is given by $j = 0.1$ and $\sigma = 0.4$).** (a–l) Sequential magnetic states colored according to the value of $n_z$. (m) Change of the total energy with time. The black symbols show the time positions corresponding to panels (a–l). (n) Dynamics rate. Notations are the same as in panel (m).

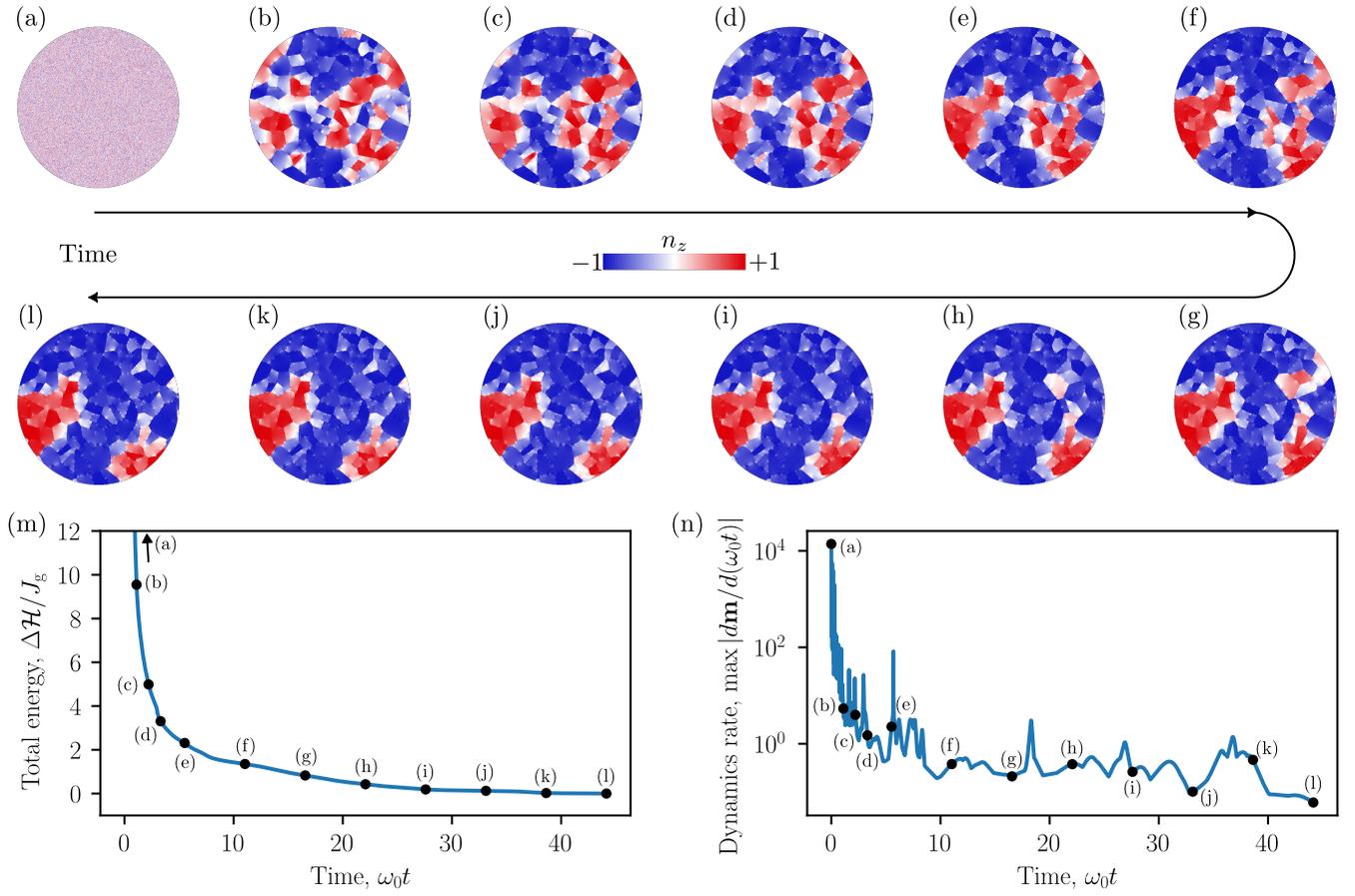

Supplementary Fig. 10. **Magnetic dynamics in ZFC-like simulations (bit diameter $12.5\ell$, grain coupling is given by $j = 0.1$ and $\sigma = 0.4$).** (a–l) Sequential magnetic states colored according to the value of $n_z$. (m) Change of the total energy with time. The black symbols show the time positions corresponding to panels (a–l). (n) Dynamics rate. Notations are the same as in panel (m).

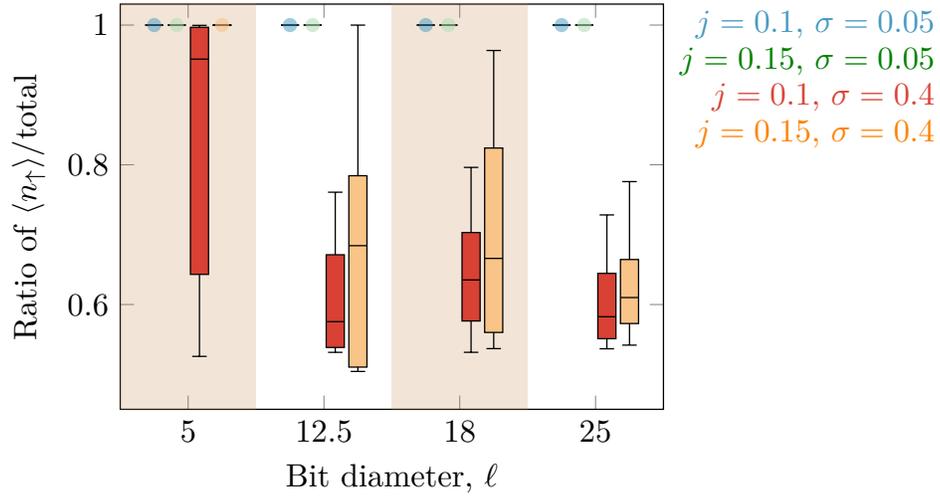

Supplementary Fig. 11. **Statistics of magnetic states in the ZFC-like procedure.** In each box, the horizontal line indicates median (c.f. with the mean value shown in Fig. 3i of the main text), box bottom and upper limits indicate lower and upper boundaries of the second and third quartile, respectively, and whiskers indicates boundaries of the first and fourth quartiles. For the certain datasets, these limits are very narrow and all together indicated by a horizontal line with a colored circle on background. For a narrow distribution of exchange bonds ($\sigma = 0.05$), bits tend to the single-domain state independently on their diameter. Growth of $\sigma$ lead to the reduction of the total area occupied by the "up" domains down to about 60% for the bit sizes considered in simulations. Statistics is collected based on 10 simulations per each set of parameters.



\end{document}